\documentclass[review]{elsarticle}

\usepackage{lineno,hyperref}
\usepackage{ulem}
\usepackage{color}
\usepackage{amsmath}
\usepackage{amssymb}

\journal{Journal of \LaTeX\ Templates}

\newcommand{\II}{I(t, {\bf r}, \oomega)}
\newcommand{\sr}{S(t, {\bf r}, \oomega)}
\newcommand{\III}{I(t, \tau, \oomega)}

\newcommand{\srt}{S(t, \tau', \oomega)}
\newcommand{\oomega}{{\bf \Omega}}

\begin{document}

\begin{frontmatter}

\title{TRINITY: a three-dimensional time-dependent radiative transfer code for in-vivo near-infrared imaging}

\author{Hidenobu Yajima$^{*}$ \fnref{myfootnote}, Makito Abe, Masayuki Umemura, Yuichi Takamizu}
\address{Center for Computational Sciences, University of Tsukuba, \\
Ten-nodai, 1-1-1 Tsukuba, Ibaraki 305-8577, Japan}
\fntext[myfootnote]{Email address: yajima@ccs.tsukuba.ac.jp}

\author{Yoko Hoshi}
\address{Preeminent Medical Photonics Education and Research Center, Hamamatsu University School of Medicine, 
1-20-1 Handayama, Higashi-ku, Hamamatsu, Shizuoka 431-3192, Japan}


\cortext[mycorrespondingauthor]{Corresponding author}


\begin{abstract}
We develop a new three-dimensional time-dependent radiative transfer code, TRINITY (Time-dependent Radiative transfer In Near-Infrared TomographY), 
for in-vivo diffuse optical tomography (DOT). 
The simulation code is based on the design of long radiation rays connecting boundaries of a computational domain, 
which allows us to calculate light propagation with little numerical diffusion.  
We parallelize the code with Message Passing Interface (MPI) using the domain decomposition technique 
and confirm the high parallelization efficiency, 
so that simulations with a spatial resolution of $\sim 1$ millimeter can be performed
in practical time. 
As a first application, we study the light propagation for a pulse collimated within $\theta \sim 15^\circ$ in a phantom, which is a uniform medium made of polyurethane mimicking biological tissue.
We show that the pulse  spreads in all forward directions over $\sim$ a few millimeters due to the multiple scattering process of photons. 
Our simulations successfully reproduce the time-resolved signals measured with eight detectors for the phantom.
We also introduce the effects of reflection and refraction at the boundary of medium with a different refractive index
and demonstrate the faster propagation of photons in an air hole that is an analogue
for the respiratory tract. 
\end{abstract}


\end{frontmatter}

\section{Introduction}

Diffuse Optical Tomography (DOT) has been proposed as a powerful tool to diagnose biological tissue, which utilizes near-infrared light at the wavelengths of $\sim 700 - 1000$ nanometer (nm) that can permeate biological tissues with the depth of $\sim 3-4$ centimeter (cm) \citep{Gibson2005, Durduran2010, Hoshi2016}.  
In the DOT, the propagation of photons in biological tissues is regulated by 
hemoglobin (Hb), water, and mitochondria. At a given wavelength in the near-infrared range, the different tissues have different optical properties like absorption and scattering coefficients, refractive index and $g$-factor.
Therefore, by measuring the photon signals escaped from surfaces, we can derive spatial information of absorption and scattering coefficients. The information 
allows us to construct the spatial map for Hb concentrations, tissue oxygen saturation ($\rm StO_{2}$), and blood flow. 
The states of them provide the diagnosis of various diseases and health conditions.  
For example, higher Hb concentrations and lower $\rm StO_{2}$ can be tracers of malignancy \citep{Corlu2007}. 

Besides, the recent development of the facility for the DOT has allowed us to detect the signal with the high time resolution, $\sim $ picoseconds \citep{Mimura2021}.  
Multiple scattering processes of photons lead to long travels in biological tissues. Photons passing through deeper inside take longer traveling time. Therefore, the time-dependent radiation transfer calculations are required to reveal the three-dimensional state in biological tissues. 
Compared to other diagnostics methods, e.g., X-ray CT (Computed Tomography) and MRI (Magnetic Resonance Imaging), the DOT has various advantages: (1) no radiative exposure, (2) non-invasive, and (3) bed-side monitoring. 
However, the DOT has faced difficulties in the reconstruction of structure, because photons propagate with complex trajectories due to multiple scattering. 
To reconstruct the structure, numerous forward models should be provided by
radiative transfer calculations \citep{Klose2002}. However, the radiative transfer in three-dimensional
space is a six-dimensional problem, and also
the radiative transfer equation (RTE) with scattering is an integro-differential equation 
that requires iteration to converge the solutions. 
 Therefore, numerical simulations take high computational costs. 
Hitherto, the effective methodology for time-dependent radiative transfer simulations has not been developed to circumvent high computational costs. 

To solve RTE, we calculate intensity that is a distribution function in multi-dimensional phase space depending on place, direction, and time. Therefore, the accurate calculations of RTE in three-dimensional space are challenging even with current computational facilities. 
Monte Carlo technique has been introduced 
as a simple approach to evaluate the radiation field stochastically
\citep{Flock1989, Boas2002, Okada2003, Fang2009, Sakakibara2016}. 
However, due to the random samplings of angle and propagation distance in this method, photon signals are contaminated with shot noises. This leads to inaccuracy in the reconstruction of biomedical states from the signals.
In order to reduce the calculation amount and memory space, the diffusion equation (DE) has been often solved instead of RTE \citep{Kannan2011}. The DE determines the propagation of radiation solely by the gradient of radiation energy density. This approximation is valid only if the radiation fields are close to isotropic   distributions due to numerous scatterings.
However, calculations based on the DE cannot keep high accuracy near surfaces or in abnormal tissues,
because the radiation fields are likely to be anisotropic there. Moreover, it is difficult for the DE to take into account the refraction and reflection at a boundary between media with different refractive indices. 
To consider the anisotropy of radiation, the approximations based on spherical harmonics with higher orders have been applied for tissue optics \citep{Aydin2004, Wright2006}. Klose et al. (2006) \cite{Klose2006} demonstrated that a simple spherical harmonics method could follow a light propagation accurately with less expensive calculation. However, as the anisotropy of the radiation field becomes higher, these methods need higher orders of the spherical harmonics, resulting in expensive calculations. 
 In addition, the hybrid method combining the DE and the RTE has been considered \citep{Tarvainen2005, Fujii2014}. Tarvainen et al. (2005) \cite{Tarvainen2005} solved the RTE only for the domain where the DE is not valid and used the DE elsewhere. Using this method, they successfully reduced calculation amounts by a factor of $\sim 2$ with keeping the accuracy. However, note that it is difficult to know the anisotropy of the radiation fields without solving the RTE in cases of highly inhomogeneous media. In such a case, the domain where the DE is invalid or a required order for the spherical harmonics is unclear.   
Therefore, to realize precise reconstruction in general cases, solving RTE is required \citep{Abdoulaev2003}. 

Recently, Fujii et al. (2018) \cite{Fujii2018} performed three-dimensional radiative transfer simulations with open multi-processing (OpenMP) parallelization of a central processing unit (CPU).
However, the spatial and the angular resolutions are limited due to the expensive calculations of RTE with a single CPU. 
To manage the larger computational amounts and memory space, utilizing multiple CPUs via Message Passing Interface (MPI) parallelization is a promising way. 
In astrophysics, three-dimensional radiative transfer simulations have been performed with multiple CPUs \cite{Nakamoto2001, Yajima2009, Yajima2011, Yajima2012h, Wise2012}. 
However, in most astrophysical simulations, RTE is solved without the time dependency, since the light crossing time is much shorter than the dynamical time of objects. 
Moreover, unlike astrophysics, we have to take into account the refraction, reflection, and highly anisotropic scattering in the DOT, which require high angular resolution. Therefore, performing time-dependent radiative transfer simulations with high resolutions for the DOT is a challenging issue. 
In this paper, we develop a novel radiative transfer code, TRINITY (Time-dependent Radiative transfer In Near-Infrared TomographY)
that is accelerated by combining MPI and OpenMP and traces light propagation with little numerical diffusion. 

\section{Radiative transfer code: TRINITY}

\subsection{Numerical procedure of solving RTE}
We calculate specific intensity $\II$ in time domain which is a distribution function in six-dimensional phase-space with time $t$, space ${\bf r}(x,y,z)$ and direction $\oomega (\theta, \phi)$.
Along a light ray, the intensity is determined by the following RTE \citep{Arridge1999}
\begin{equation}
\frac{1}{c({\bf r})}\frac{\partial \II}{\partial t} + \oomega \cdot \nabla \II = - \left[ \mu_{\rm a}({\bf r}) + \mu_{\rm s}({\bf r}) \right] \II + \mu_{\rm s}({\bf r}) \int_{4 \pi}  I(t, {\bf r}, \oomega') p({\bf r}, \oomega', \oomega) d\oomega',
\label{eq:RTE}
\end{equation}
where $c ({\bf r})$ is the light speed at position $\bf r$, 
$\mu_{\rm a}(\bf r)$ and $\mu_{\rm s} (\bf r)$ are the absorption and scattering coefficients, $p({\bf r}, \oomega', \oomega)$ is a phase function from incoming direction $\oomega'$ to outward direction $\oomega$ in a scattering process and it is normalized to be unity to the angle integral. The light speed  changes with the refractive index $n$ as $c = c_{0}/n$ where $c_{0}$ is the light speed in vacuum. 
Most previous studies solving the RTE have discretized the above equation and followed the light propagation by integrating it \citep{Fujii2017}. 

In this study, we consider a different form of the RTE. 
First, we introduce source function defined by
 \begin{equation}
\sr = \frac{\mu_{\rm s}({\bf r})}{\mu_{\rm a}({\bf r}) + \mu_{\rm s}({\bf r})} \int p({\bf r}, \oomega', \oomega) I(t, {\bf r}, \oomega') d\oomega'.
\end{equation}
%
If the time derivative term is larger than the space derivative one
in the left-hand side of Equation (\ref{eq:RTE}), 
the equation can be approximated to be 
\begin{equation}
\frac{d \II}{d \tau_t }  = - \II + \sr,
\end{equation}
where $d\tau_t \equiv \left[ \mu_{\rm a}({\bf r}) + \mu_{\rm s}({\bf r}) \right] c({\bf r}) dt$. 
The formal solution of this equation is
\begin{equation}
I(t, {\bf r}, \oomega) = I(0, {\bf r}, \oomega)~{\rm exp} (- \tau_t) + \int_{0}^{\tau_t} S(t', {\bf r}, \oomega) {\rm exp}(\tau_t' - \tau_t) d\tau_t'.
\end{equation}
On the other hand, if the space derivative term dominates the time derivative one is,
Equation (\ref{eq:RTE}) can be approximated to be 
\begin{equation}
\frac{d \II}{d \tau}  = - \II + \sr,
\label{eq:RTE2}
\end{equation}
where we introduce the optical depth ($\tau$) defined by $d\tau = \left[ \mu_{\rm a}({\bf r}) + \mu_{\rm s}({\bf r}) \right] dr$.
The formal solution of this equation is
\begin{equation}
I(t, \tau, \oomega) = I(t, 0, \oomega)~{\rm exp} (- \tau) + \int_{0}^{\tau} S(t, \tau', \oomega) {\rm exp}(\tau' - \tau) d\tau'.
\end{equation}
To incorporate the effect of time derivative into this formal solution, 
we employ a retarded form as
\begin{equation}
I(t, \tau, \oomega) = I(0, 0, \oomega)~{\rm exp} (- \tau) + \int_{0}^{\tau} S(t, \tau', \oomega) {\rm exp}(\tau' - \tau) d\tau'.
\label{eq:RTE3}
\end{equation}

We solve this equation at each time step with  TRINITY, 
which corresponds to the integration of  Equation (\ref{eq:RTE}). 
We evaluate the source function in the integration by linear interpolation between two points as
\begin{equation}
\srt = \left( 1 - \frac{\tau'}{\tau}\right) S_{0}
+ \frac{\tau'}{\tau} S_{1}
\end{equation}
where $S_{0} = S(0, 0, \oomega)$ and $S_{1} = S(t, \tau, \oomega)$. 
Then, we discretize equation (\ref{eq:RTE3}) as
\begin{equation}
\begin{split}
\III = & I_{0}~{\rm exp} (-  \tau)   \\
&+ S_{0}\frac{1 - (1+ \tau){\rm exp}(- \tau)}{ \tau} 
+ S_{1}\frac{ \tau -1 + {\rm exp}(- \tau)}{ \tau},
\end{split}
\end{equation}
where $I_{0} \equiv I(0, 0, \oomega)$.

\begin{figure}
	\begin{center}
		\includegraphics[width=8.0 cm,clip]{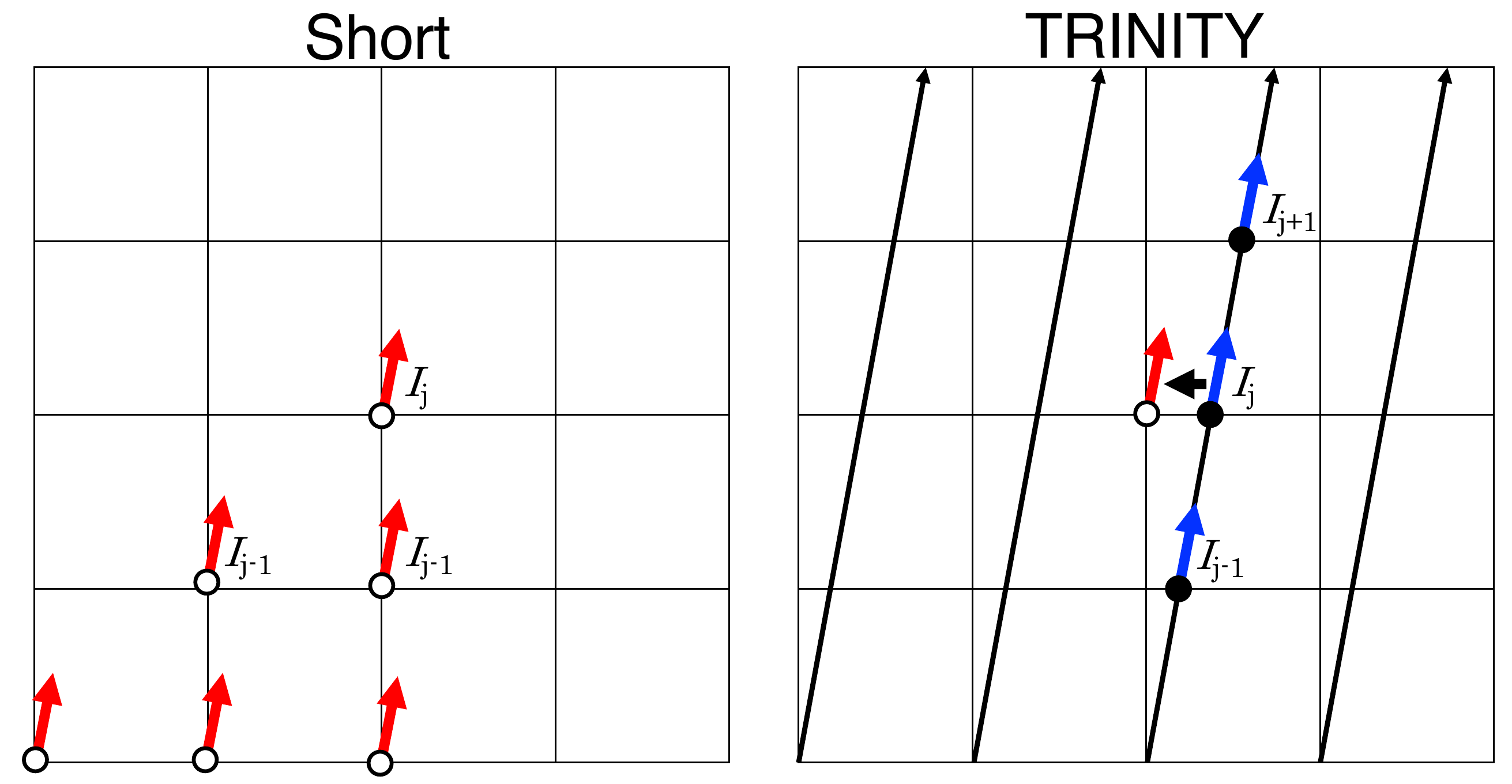}
	\end{center}
	\vspace{-5mm}
	\caption{
	Schematic view of TRINITY and short-characteristic method. Spatial grids distribute at the cross points of the lattice. 
	Radiative rays are designed independently to the spatial grids in TRINITY. The intensity at a specific spatial grid 
	is interpolated from a nearest radiative ray. 
		}  
	\label{fig:art}
\end{figure}

In TRINITY, we assign light rays independently of spatial grids as shown in Figure~\ref{fig:art}. This basic idea was proposed in our previous works to calculate radiation transfer in galactic or metagalactic space \citep{Yajima2009, Yajima2011, Yajima2012h}. 
We evaluate $\II$ at points where a ray crosses spatial grids by using the information of an up-wind grid as
\begin{equation}
\begin{split}
\II = & I(t - \frac{\Delta r}{c} , {\bf r} - \Delta {\bf r}, \oomega)~{\rm exp} (- \Delta \tau)   \\
&+ S(t - \frac{\Delta r}{c} , {\bf r} - \Delta {\bf r}, \oomega) \frac{1 - (1+\Delta \tau){\rm exp}(-\Delta \tau)}{\Delta \tau} 
\\
&+ \sr \frac{\Delta \tau -1 + {\rm exp}(-\Delta \tau)}{\Delta \tau}, 
\end{split}
\end{equation}
where $\Delta r$ and $\Delta \tau$ are distance and optical depth between the two radiation grids.
The absorption, scattering coefficients, and refractive index on a crossing point are interpolated from those on nearest spatial grids. 
In most previous works solving RTE, a light ray is set at a spatial grid and not connected with a ray of an upwind grid, which is called the short-characteristic method \citep{Stone1992} or the finite-element method \citep{Arridge1999}. This method is subject to considerable numerical diffusion as a result of multiple interpolations as shown in Section 2.2.  
On the other hand, TRINITY is based on a long ray connecting boundaries of a calculation box. 
In this scheme, the intensity at a specific spatial grid is evaluated by only one interpolation from nearest rays. Thus,  numerical diffusion is substantially reduced. 

The phase function for the scattering processes in biological tissue is generally modeled by 
the Henyey-Greenstein function:
\begin{equation}
\phi(\oomega \cdot \oomega') = \frac{1}{4 \pi} \frac{1 - g^{2}}{(1 + g^{2} - 2g\oomega \cdot \oomega')^{\frac{3}{2}}},
\end{equation}
where $g$ is an anisotropy factor. For example, $g=0$ represents isotropic scattering and $g=1$ does complete forward scattering. In cases of biological tissues, $g$ is usually larger than $\sim 0.8$ \citep{Cheong1990, Jacques2013}. Therefore, the photons propagate with highly forward scattering processes. 

For the discretization of angle, we apply HEALPix \citep{Gorski2005} that was developed for analyzing the cosmic microwave background. HEALPix divides a spherical surface with square pieces of equal area and can change the resolution hierarchically. As fiducial simulations, we settle 3072 angle bins for radiative transfer (RT) simulations.  
To estimate the source function, we sum up intensities for the discretized angles as
\begin{equation}
S(t, {\bf r}, {\bf \Omega}) = \frac{\mu_{\rm s}({\bf r})}{\mu_{\rm a}({\bf r}) + \mu_{\rm s}({\bf r})} \sum_{j=1}^{N'_{{\bf \Omega}}} p({\bf r}, {\bf \Omega}_{j}, {\bf \Omega}) I(t, {\bf r}, {\bf \Omega}_{j}) \Delta {\bf \Omega},
\end{equation}
where $N'_{\oomega}$ is the number of angle bins.
The calculation amount of evaluating $\sr$ to all directions increases in proportion to $N^{'2}_{\oomega}$.
Thanks to HEALPix's algorithm, we simply change the level of the hierarchical structure. 
In this work, we set $N'_{\oomega}=3072/16 = 192$ 
by coarse-graining of $3072$ finer angular levels. 
So, the mean of $\II$ at 16 finer angles  is used for the calculation of $S$. 

The evaluation of $\sr$ requires the information of $\II$ at the current time step and vise versa. 
To obtain both $\II$ and $\sr$ consistently, we iterate RT simulations until both are converged at each time step. In the iterations, we first set the source function at a previous time step $S(t-\Delta t, {\bf r}, \oomega)$ as an initial dummy.

At the boundary of media with different refractive indices, we consider refraction and reflection. 
The refraction angle is estimated based on the Snell's law as
\begin{equation}
\frac{n_1}{n_2} = \frac{{\rm sin} \theta_2}{{\rm sin} \theta_1},
\label{eq:refraction}
\end{equation}
where $n_1$ and $n_2$ are refractive indices. 
By using the refraction angle, the reflection rate is calculated by the Fresnel's law,
\begin{equation}
R(n/n_0,\theta) =  \begin{cases}
\frac{1}{2} \left[  \frac{{\rm sin}^{2}(\theta_1 - \theta_2)}{{\rm sin}^{2}(\theta_1 + \theta_2)}
 + \frac{{\rm tan}^{2} (\theta_1 - \theta_2)}{{\rm tan}^{2}(\theta_1 + \theta_2)} \right], 
 ~~~~ \theta < \theta_{\rm c} \\
1, ~~~~~~~~~~~~~~~~~~~~~~~~~~~~~~~~~~~~~~~~\theta \ge \theta_{\rm c}
\end{cases}
\label{eq:reflection}
\end{equation}
where $\theta_{\rm c}$ is the critical angle. 

\subsection{Test calculations with TRINITY}
\begin{figure}
	\begin{center}
		\includegraphics[width=7.0 cm,clip]{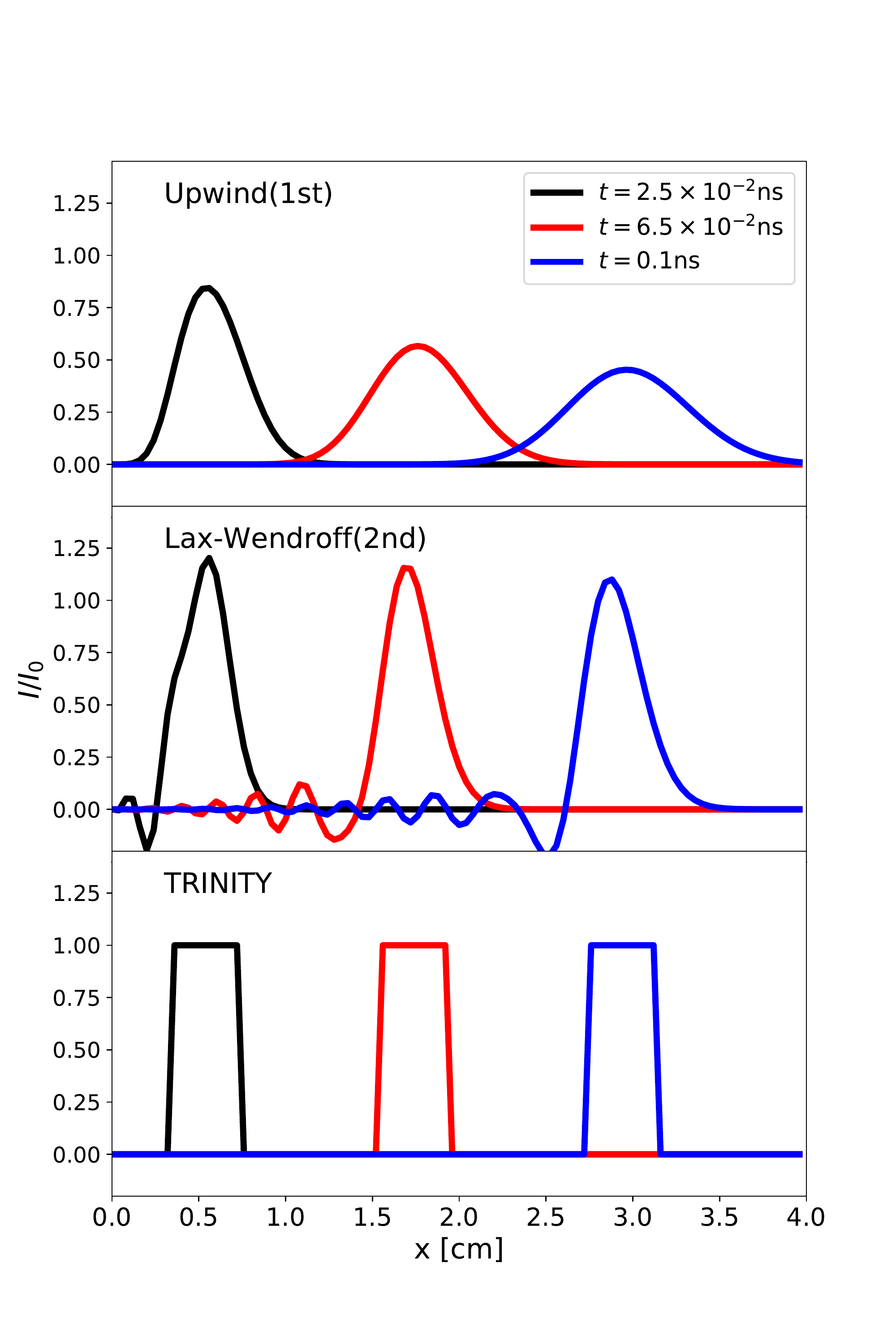}
	\end{center}
	\vspace{-5mm}
	\caption{
	One-dimensional pulse simulations. The pulse is injected for $1.3\times10^{-2}$ ns. 
	Solid lines show that spatial distributions of photon energy with different time steps: $t=2.5\times10^{-2}$ ns (black), $6.5\times10^{-2}$ ns (red), and $0.1$ ns (blue). Upper, middle and lower panels show 1st order upwind, Lax-Wendroff (2nd order), and TRINITY (this work). 
	}  
	\label{fig:1d_scheme}
\end{figure}

As a first test, we calculate the propagation of a pulse in one-dimensional space 
without absorption and scattering, i.e., light propagation in a vacuum. 
Figure~\ref{fig:1d_scheme} shows the spatial distributions of photon energy densities.
Here, we inject a pulse with duration of $1.3\times10^{-2}$ nanosecond (ns) from $x=0$ cm. The calculation box size is $4~\rm cm$ and the number of spatial grids is set to be 100. 
The time resolution should be higher than the light crossing time over two grids. We choose the time resolution as $\frac{1}{10} \times$ the light-crossing time, i.e., $\Delta t = \frac{1}{10} \times \frac{\Delta x}{ {\rm c}} = 1.3 \times10^{-4} ~\rm ns$, where $\Delta x$ is the length between grids. This is corresponding to a courant number 0.1. 
In comparison, we also solve Equation~(\ref{eq:RTE}) by discretizing with the upwind scheme (1st order)  and Lax-Wendroff (2nd order ) schemes \citep{Press1992}. 
As shown in Figure~\ref{fig:1d_scheme}, the upwind scheme results in artificial extension of  energy distributions due to numerical diffusion
according as the pulse propagates. The peak energy decreases as the propagation distance increases and becomes $\sim 50\%$ at $x \sim 3.0~\rm cm$.
In the case of the Lax-Wendroff, the numerical diffusion is suppressed, but the rectangular shape of the beam is distorted and also oscillations emerge. 
We find that TRINITY preserves the shape of the pulse. 
Both numerical diffusion and oscillations are significantly suppressed. 
Note that, in the case of TRINITY, the data sizes of $I$ and $S$ are larger than those in previous works because of retaining the information of $10$ time bins.

\begin{figure}
	\begin{center}
		\includegraphics[width=10.0 cm,clip]{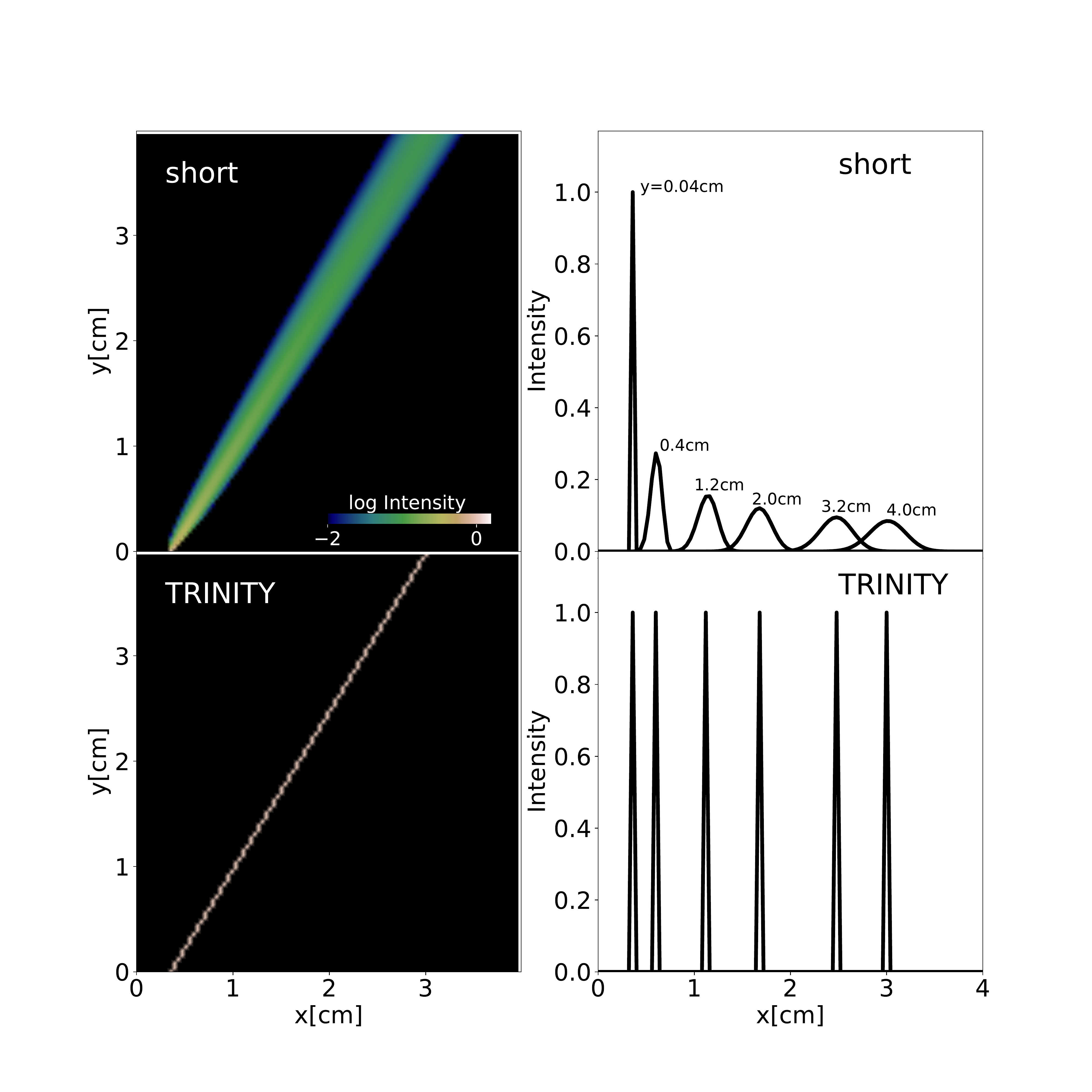}
	\end{center}
	\vspace{-5mm}
	\caption{
	Left panels: Propagation of beam in two-dimensional space.  Upper panel shows the result of short-characteristic method and lower panel does TRINITY. The color represents intensity at the opening angle $\theta=56.3^\circ$ from  $x$-axis. 
	Light panels: Distribution of energy density along x-axis at specific $y$ positions. Different lines represent different $y$ positions. 
	}  
	\label{fig:2d_comp}
\end{figure}

Next, we compare TRINITY with the short-characteristic method in the case of propagation of collimated beam in two-dimensional space. 
Figure~\ref{fig:2d_comp} presents the spatial distributions of intensity in the direction of beam. 
The calculation box is set as $4.0 ~{\rm cm}~\times~ 4.0 ~{\rm cm}$ with the spatial grids of $101 \times 101$. 
To clarify the difference between schemes, we investigate the propagation of the beam in vacuum ($\mu_{\rm a} = \mu_{\rm s} = 0$). The beam is successively injected from $x=0.4~{\rm cm}$ and $y = 0.0~\rm cm$ with the inclination angle $\theta = 56.3^{\circ}$. 

In the case of the short-characteristic method, the beam spreads due to the repeated interpolations from up-wind grids according as it propagates . In contrast, TRINITY preserves the collimated shape. 
Right panels of Figure~\ref{fig:2d_comp} represent the intensity distributions along $x$-axis at specific $y$ positions. 
In the case of the short-characteristic method, the peak values decrease as the $y$ position increases. 
At the opposite boundary, the peak value is reduced to $8.5 \times 10^{-2}$ 
from unity as the original value.
On the other hand, TRINITY keeps the peak value of $\sim 1$ even far from the injected position. 

\begin{figure}
	\begin{center}
		\includegraphics[width=10.0 cm,clip]{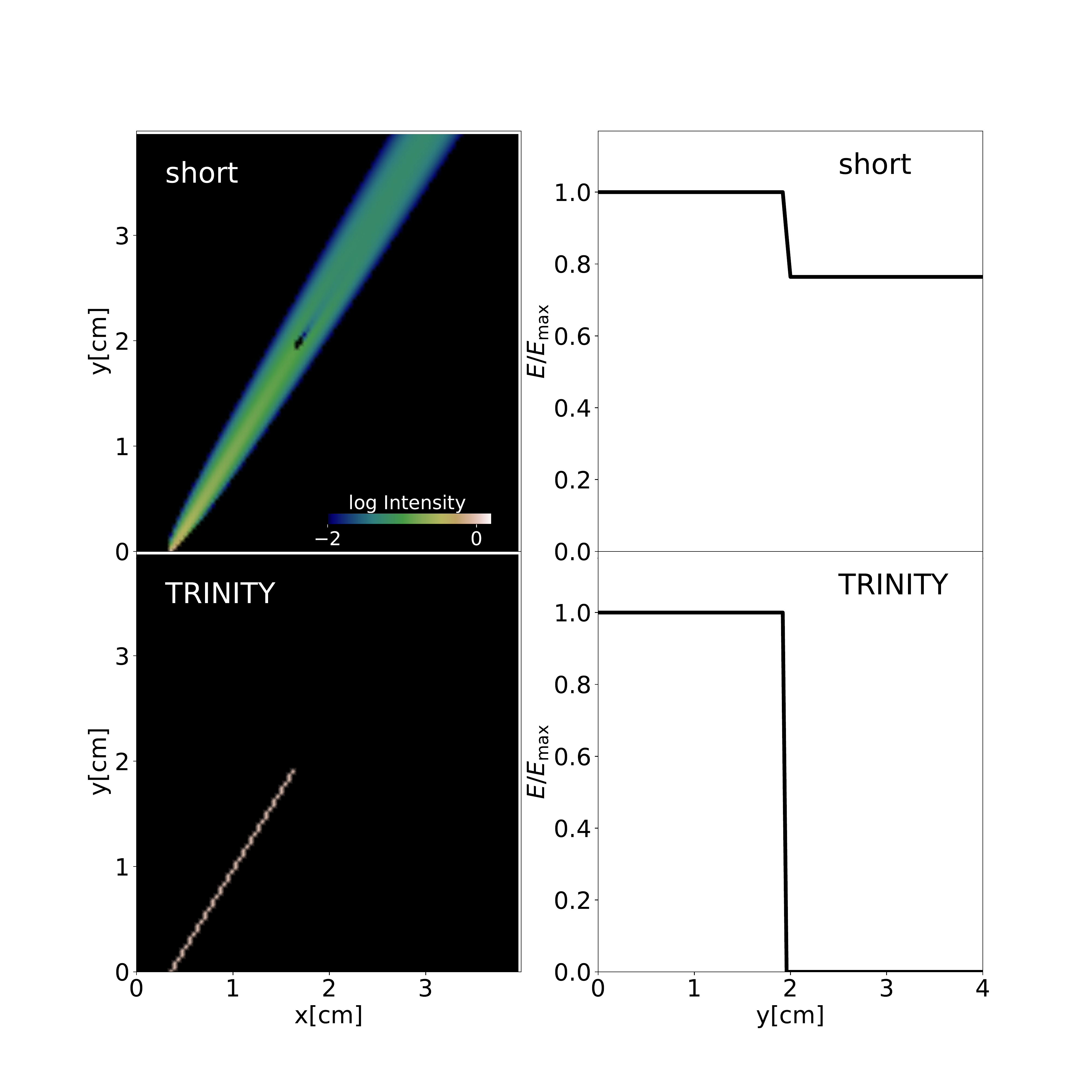}
	\end{center}
	\vspace{-5mm}
	\caption{
	Left panels: Same as Figure~\ref{fig:2d_comp}, but a grid with a high absorption efficiency is set at $x=1.7$ and $y=2.0~\rm cm$. 
	Light panels: The photon energy density  integrated along $x$-axis normalized by the maximum value as a function of $y$ position.
	}  
	\label{fig:2d_comp_abs}
\end{figure}

In addition, we demonstrate the effect of the numerical diffusion for a case with an absorbing area. 
In addition to the above situation, we place a grid with a high absorption coefficient on the way of the beam at $x=1.7$ and $y=2.0~\rm cm$. Figure~\ref{fig:2d_comp_abs} presents the spatial distribution of intensity. In the case of TRINITY, the beam completely disappears at the absorption grid,  whereas in the case of the short-characteristic method, the numerical diffusion allows photons to propagate behind the absorption grid. The figure shows that the shadowing area behind the absorption grid is gradually filled as the light propagates. Right panels present the radiation energy density integrated along $x$-axis as a function of $y$ position. We find that $76\%$ of the beam energy avoids the absorption grid and reaches the boundary in the case of the short-characteristic method. 
Therefore, we suggest that the numerical diffusion can induce fake signal at a detector on surface if there is an abnormal area smaller than the width of the numerical diffusion. 

\begin{figure}
	\begin{center}
		\includegraphics[width=10.0 cm,clip]{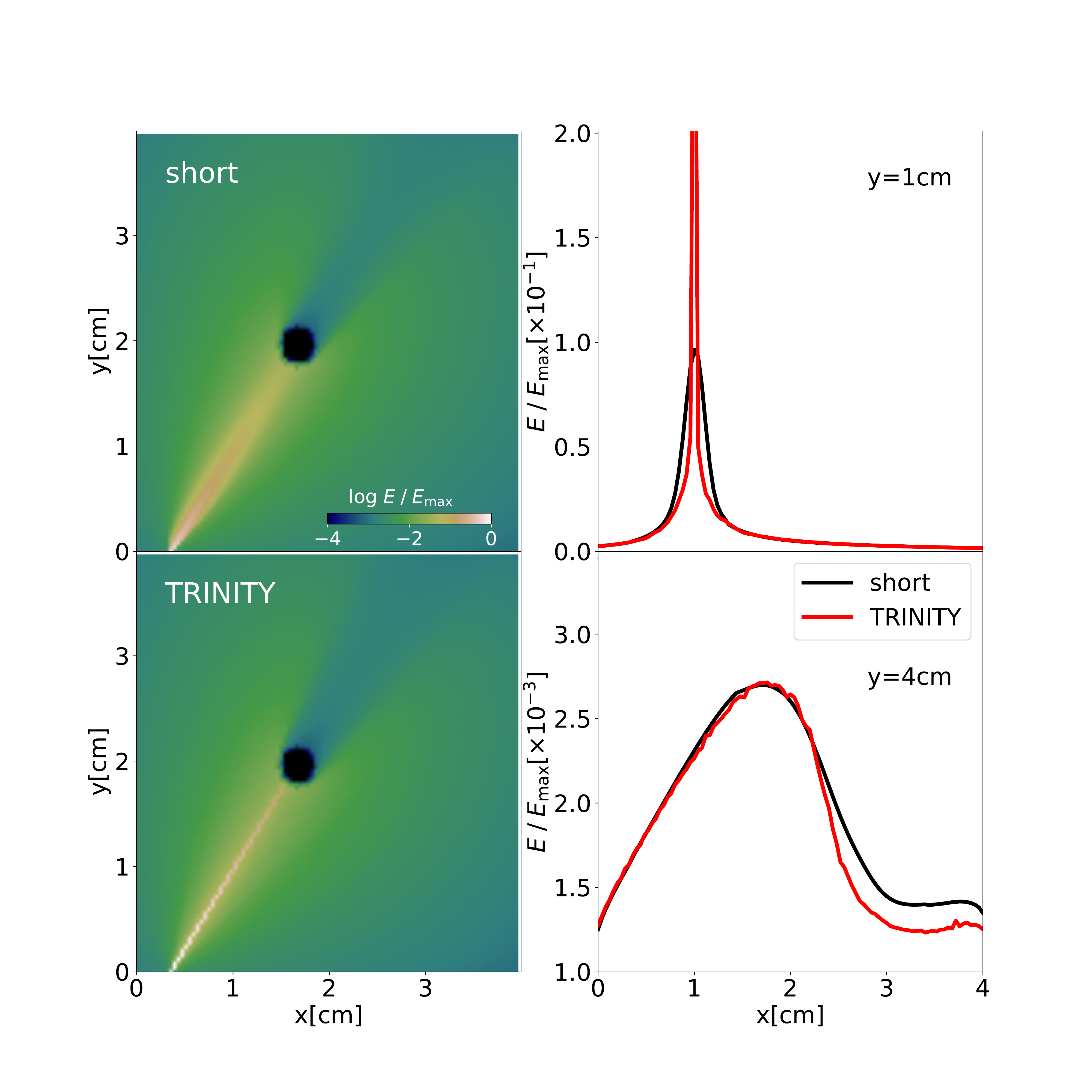}
	\end{center}
	\vspace{-5mm}
	\caption{
	Left panels: Propagation of beam in two-dimensional space with absorption and scattering processes. The area at $x=1.7$ and $y=2.0~\rm cm$ with a radius of $0.2~\rm cm$ has absorption coefficient of $100.0 ~\rm cm^{-1}$. The scattering coefficient of $1.0 ~\rm cm^{-1}$ is set uniformly. The color shows photon energy density normalized by the maximum value ($E_{\rm max}$) in the case of TRINITY. The snapshots at $t=0.27~\rm ns$ are used. 
	Light panels: The energy density distributions along $x$-axis at $y=1.0$ and $4.0~\rm cm$ (boundary). Black and red lines represent the short-characteristic and TRINITY methods, respectively. 
	}
	\label{fig:2d_comp_scalow}
\end{figure}

In cases with scattering processes, the difference between the schemes is likely to be smaller. 
Here, we also investigate the beam propagation in a scattering medium. 
Figure~\ref{fig:2d_comp_scalow} shows two-dimensional beam tests with the scattering. Here, we set the scattering coefficient as $\mu_{\rm s}=1.0~\rm cm^{-1}$ uniformly and put an absorption area at $x=1.7$ and $y=2.0$ cm with a radius of $0.2$ cm. The absorption efficiency is set as $\mu_{\rm a}= 100.0~\rm cm^{-1}$. 
The number of angle bins is 384, i.e., $\Delta \theta = 2 \pi / 384$. 
Left panels present two-dimensional maps of photon energy density normalized by the maximum value in the case of TRINITY. The snapshots at $t=0.27~\rm ns$ are used. 
Unlike the above simulations in a vacuum, the beam spreads out via the scattering processes as it propagates. 
Therefore, even the simulation with TRINITY shows that the light propagates beyond the absorption area. 
The difference between the schemes does not look significant. This indicates that the radiation field is dominated by the physical diffusion due to the scattering rather than the numerical diffusion.  
Right panels show the spatial distributions of the energy densities along $x$-axis at $y=1.0$ and $4.0$ cm. At $y=1.0$ cm, the beam in the case of the short-characteristic method is more extended due to the numerical diffusion. This makes the peak value lower than the case of TRINITY by a factor of $\sim 4$. At $y=4.0$ cm, the energy density of TRINITY is lower than that of the short-characteristic method at $x \sim 2.5 - 4.0$ where is the shadow area. The maximum difference is $18~\%$ at $x=2.7$ cm. In the case of the short-characteristic method, the beam spreads out more, resulting in lower photon absorption. 
Also, the area behind the absorption is recovered more efficiently because of the numerical diffusion. 
Thus, the energy density at the shadow area in the short-characteristic method is higher than that in the case of TRINITY. 

\subsection{Parallelization}

\begin{figure}
	\begin{center}
		\includegraphics[width=6.0 cm,clip]{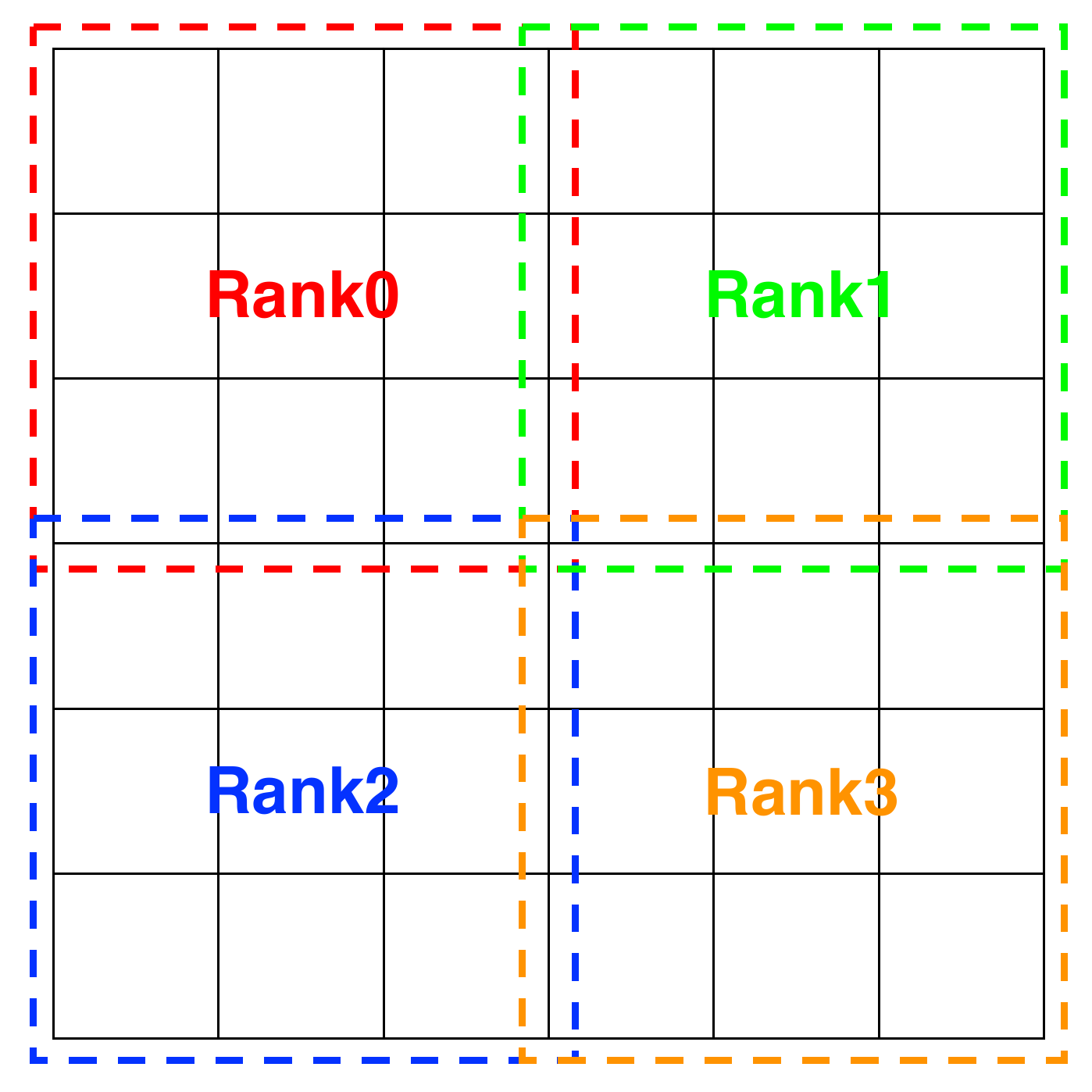}
	\end{center}
	\vspace{-5mm}
	\caption{
	Schematic view of MPI parallelization with the domain decomposition in TRINITY.	}  
	\label{fig:parra}
\end{figure}

Three-dimensional numerical simulations of RTE require huge calculation amounts and memory. 
To overcome the difficulty, we combine MPI and openMP parallelization techniques. 
To implement MPI, we decompose the calculation domain as shown in Figure~\ref{fig:parra}. 
Each CPU rank calculates RTE in its own domain. The size of each domain is set to $L / (N_{\rm CPU})^{1/3}$, where $L$ is the size on a side of the entire calculation box and $N_{\rm CPU}$ is the number of CPUs. 
For the loop-calculations of the angle bins, we accelerate the code with openMP using the cores of each CPU. 
Then, before proceeding to the next time step, each domain communicates with adjacent domains and gets intensities at boundary grids.  
As a result of such hybrid parallelization, we can utilize more than 100 CPU cores simultaneously. 
This allows us to handle direct simulations of RTE in three-dimensional space with more than 100 grids on one side.

We confirm that calculations are accelerated efficiently by increasing the number of CPU core.
However, for the cases with a small number of spatial grids (e.g., $\lesssim 40^{3}$), the parallelization efficiency decreases even if a lot of CPUs are utilized. This is because the data communication of intensities at boundary grids between the domains becomes a bottleneck. 
In our simulations, we use $\gtrsim 100^{3}$ grids to achieve the resolution of $\lesssim 1$ millimeter to parts of a human body with the size of $\lesssim 10~\rm cm$. 

\subsection{Calculation steps}
We summarize the calculation steps in TRINITY. 

Step 1: Assigning  light rays connecting from a boundary to another side boundary. The total number of rays is $\sim N_{\rm ang} \times N_{\rm grid}^{2}$, where $N_{\rm ang}$ is the number of angle bins and   $N_{\rm grid}$ is the number of spatial grid on one side. 

Step 2: Decomposing a calculation box. 

Step 3: Solving RTE on each radiative ray.   

Step 4: Data communication of intensities at boundaries between adjacent domains.

Step 5: Calculating the source function. 

Step 6: Go to step 3) until the source function becomes consistent with that derived from the intensities at the current time step.

Step 7: Go to the next time step if the intensity and source function are consistent.  

%
%

\section{Results}

\subsection{Application to a phantom}
\label{sec:phantom}

\begin{figure}
	\begin{center}
		\includegraphics[width=12.0 cm,clip]{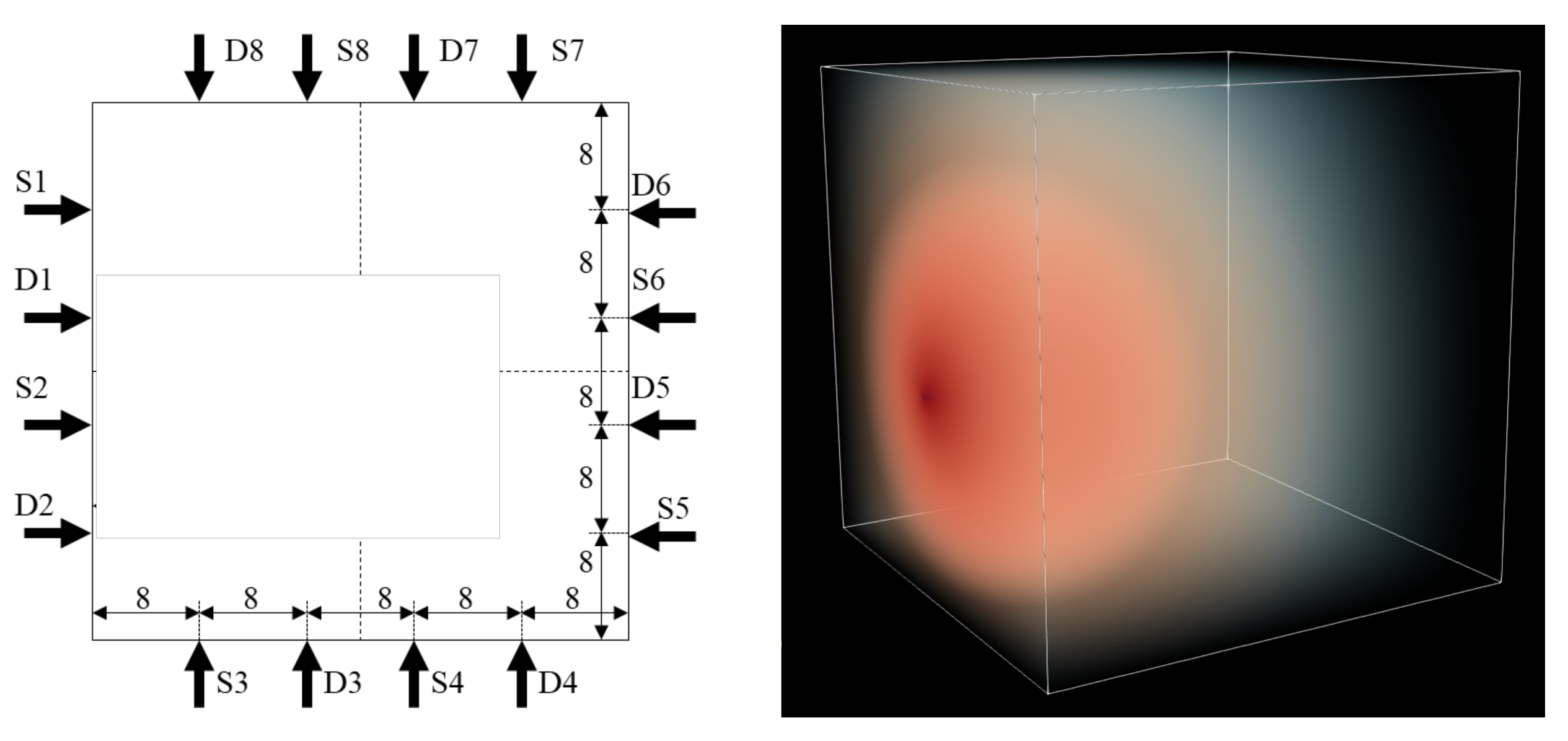}
	\end{center}
	\vspace{-5mm}
	\caption{
	Left panel: Schematic view of a pulse experiment for phantom. S1-S8 show positions of pulse injection. D1-D8 represent the positions of detectors.
	Right panel: radiation field obtained from the radiative transfer simulation. The color shows the photon energy density in log-scale.	}  
	\label{fig:phantom_3d}
\end{figure}

As applications of TRINITY, we model the light propagation in a phantom made of polyurethane which mimics biological tissues.  The properties related to RT are summarized in Table~\ref{table:phantom}.
In the left panel of Figure~\ref{fig:phantom_3d}, 
the schematic picture of the phantom is shown.  
The simulation box has the size of $(4.0\;{\rm cm} \times 4.0\;{\rm cm} \times 4.0\;{\rm cm})$ and is discretized with grids of $101^{3}$. 
S1-8 and D1-8 represent the candidate positions of source and detectors. These positions are on the same $x-y$ plane at $z=2.0~\rm cm$.
We inject a pulse from S2 position and detect signals at D1-8 positions. 
The pulse is highly collimated within $\theta = 15^{\circ}$. The opening angle of the detectors is also the same. 
The time resolution is set to be $\frac{1}{5} \times$ the light-crossing time between grids, i.e., $\Delta t = \frac{1}{5} \times \frac{\Delta x}{({\rm c}/n) }= 4.0 \times 10^{-4} ~\rm ns$.
We pursue the light propagation for $3.5~\rm ns$ after the pulse is injected.
The resultant total number of time steps is $8680$. 
In this simulation, we assume that all photons are absorbed once they reach the boundary, i.e., there is no reflection. In practice, the phantom is covered by black carbon as the photo-absorption media in the experiments compared with our simulations. 

\begin{figure}
	\begin{center}
		\includegraphics[width=10.0 cm,clip]{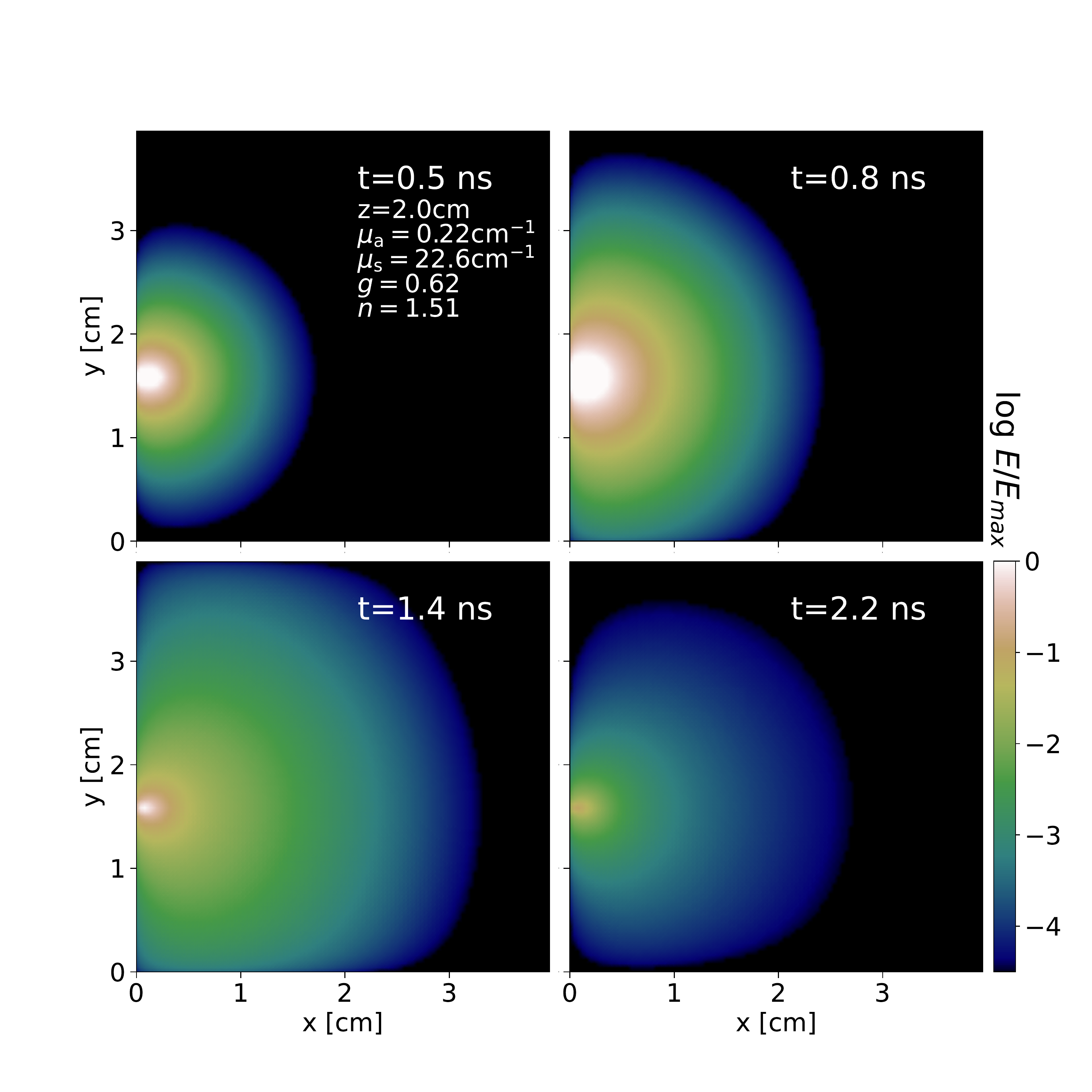}
	\end{center}
	\vspace{-5mm}
	\caption{
	$x-y$ two-dimensional map of photon energy density at $z=2.0$ cm in log-scale.
	The values are normalized by the energy peak at $t=1.4~\rm ns$.
	}  
	\label{fig:2dmap}
\end{figure}

The right panel of Figure~\ref{fig:phantom_3d} shows the snapshot of the three-dimensional energy distributions at  $t = 0.8$ ns. Because of the multiple scattering processes, the radiation fields spread over the box and the energy tends to be distributed spherically. 
Figure~\ref{fig:2dmap} presents the energy distributions on $x-y$ plane at $z = 2.0~\rm cm$. 
Despite the narrow beam pulse, the radiation fields widely spread even near the injection point. 
Because of the high scattering coefficient, the trajectories of photons are bent typically with the mean free path $\frac{1}{\mu_{\rm s} (1 - g)} = 7.1 \times 10^{-2} ~\rm cm$. If there is no absorption, the typical number of scatterings in the box is $\sim (\tau (1-g))^{2} \sim 3.2 \times 10^{3} $.  Thus, the box is filled with photons eventually. 
In addition, even after the stop of injection, 
an appreciable number of photons are trapped due to multiple scattering processes for longer time than the light crossing time, $t_{\rm cross} = 4.0 / {\rm c} / n = 8.8\times10^{-2}~\rm ns$, as seen in the bottom right panel. 
Thereafter, most photons gradually disappear from the box due to the absorption or escape. 

\begin{figure}
	\begin{center}
		\includegraphics[width=11.0 cm,clip]{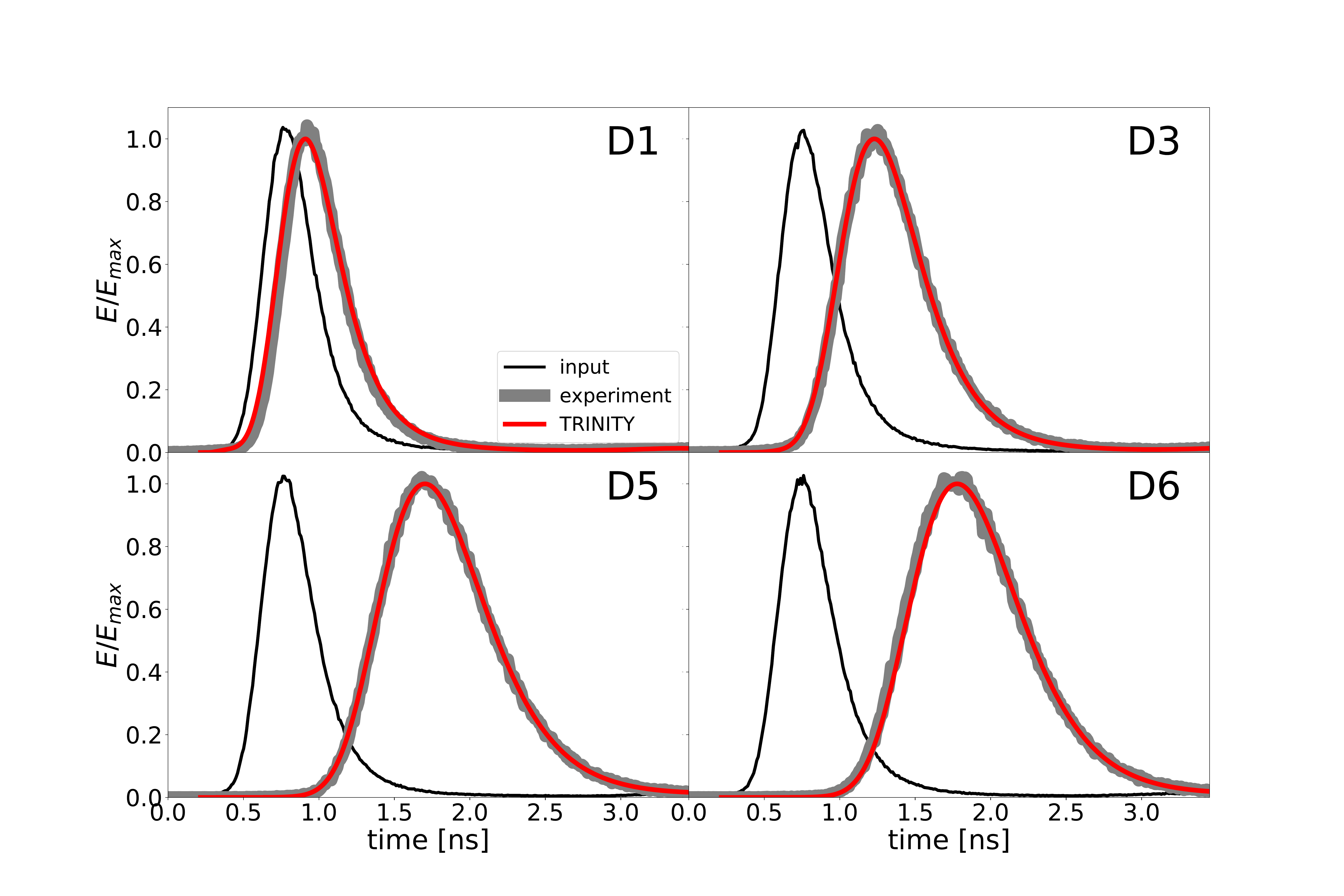}
	\end{center}
	\vspace{-5mm}
	\caption{
	Time-resolved light signal. Each panel shows the light signal taken from different detectors as shown in Figure~\ref{fig:phantom_3d}. Black solid lines represent the injected pulse. Thick gray and red curves show the experimental data and the simulation results. 
	} 
	\label{fig:phantom}
\end{figure}

Figure~\ref{fig:phantom} presents the comparison of simulation results (red lines)
with the experimental data (gray lines). The experimental signals show slight fluctuations due to the noise associated with the instrument.
We find that the shape of the output signal nicely matches the experimental data. 
For the detectors further from the injected position, 
the energy peaks arrive at later time. 
In addition, the full width at half maximum (FWHM) increases with the distance from S2. 
For example, FWHM for D5 is $\Delta t = 1.8~\rm ns$ and $0.8~\rm ns$ for D1.
This means that there is wider variation of photon trajectories (i.e., traveling time) in the case of D5. 

\begin{table*}
\begin{center}
\begin{tabular}{ccccccccc}
\hline
Properties of phantom&   \\
\hline
Box size & 4.0 cm \\
$\mu_{\rm a}$ & 0.22 cm$^{-1}$\\
$\mu_{\rm s}$ & 22.6 cm$^{-1}$\\
$g$ & 0.62 \\
$n$ & 1.51 \\
\hline
\end{tabular}
\caption{
Properties of phantom made by polyurethane:  $\mu_{\rm a}$ and $\mu_{\rm s}$ 
are absorption and scattering coefficients, $g$ is an anisotropy factor, and $n$ is the refractive index. 
}
\label{table:phantom}
\end{center}
\end{table*}

\subsection{Phantom test with an absorber}
If there are abnormal parts like cancer or brain ischemia for instance, 
the density of hemoglobin or the redox state of cytochrome c oxidase in mitochondria can become different from that in normal tissue. 
Therefore, the absorption coefficient changes locally, resulting in the different signals at the surface.
Here, we test the cases with embedded abnormal media of cylinder shape over $z$-direction with radius $0.25~\rm cm$ in the cross-section. The abnormal medium is located at $x=1.6~\rm cm$ and $y=2.0~\rm cm$.
The basic setup is the same as the simulation in Sec. \ref{sec:phantom}.
The absorption coefficient in the abnormal medium is enlarged by a factor of 3 (a), 10 (b), 30 (c), or 100 (d) from the background tissue. 

\begin{figure}
	\begin{center}
		\includegraphics[width=10.0 cm,clip]{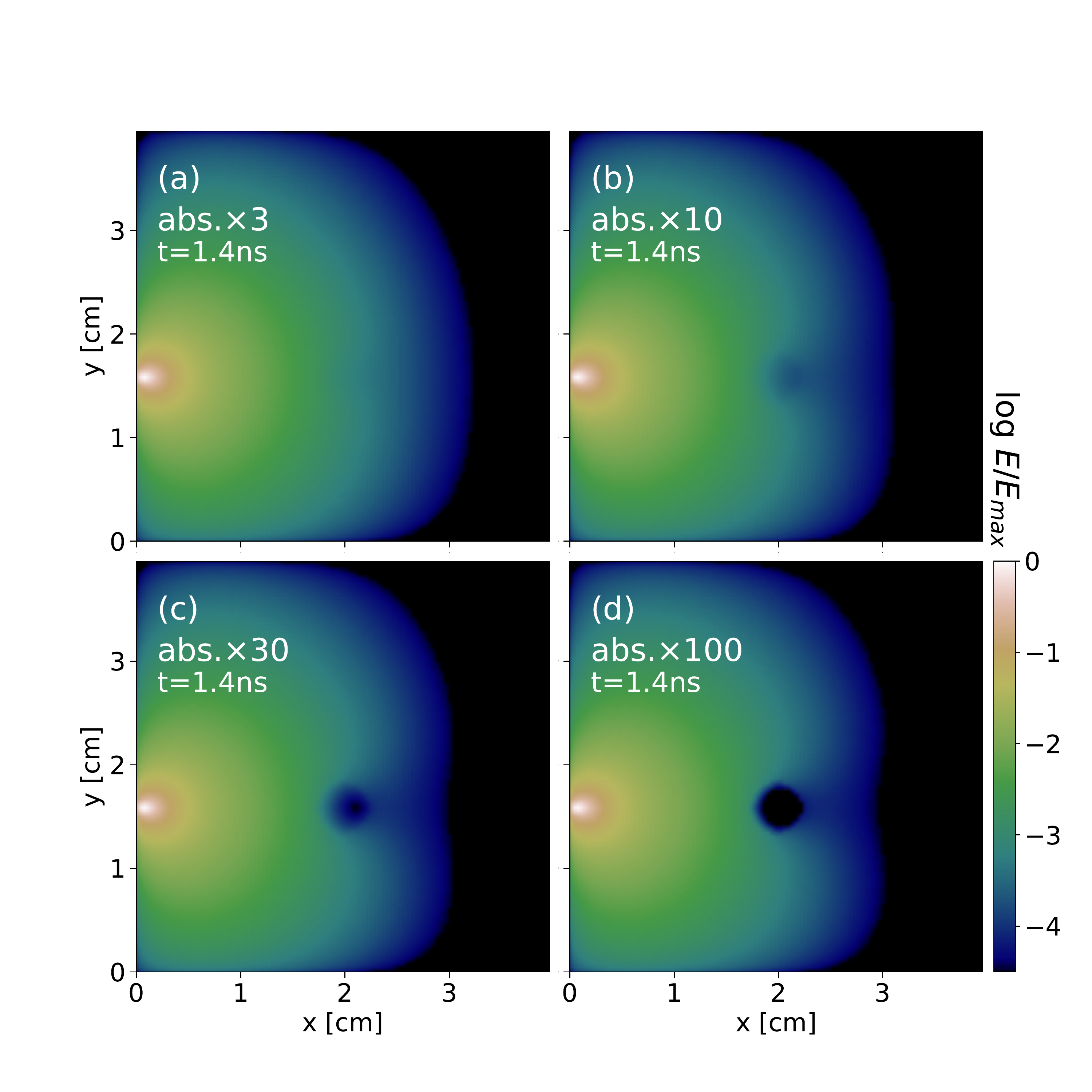}
	\end{center}
	\vspace{-5mm}
	\caption{
	Same as Figure~\ref{fig:2dmap} for snapshots at $t=1.4~\rm ns$, but with stronger absorption media at $x=2.0$ cm and $y=1.6$ cm with a radius of $0.25~\rm cm$. The absorption efficiencies are increased from background media by a factor of 3 (panel (a)), 10 (panel (b)), 30 (panel (c)), or 100 (panel (d)). 
	The energy density is normalized by the peak value of panel (a).
	}
	\label{fig:2dmapwabs}
\end{figure}

Figure~\ref{fig:2dmapwabs} shows the distributions of the photon energy density at $t = 1.4~\rm ns$.
As the absorption coefficient increases, the energy densities in the abnormal areas decrease. 
In the case (d), the optical depth of the abnormal medium is $\tau = 22.4$, and most of photons are absorbed at the depth of $\sim 1/\mu_{\rm a} = 4.5 \times 10^{-2}~\rm cm$. 
Thus, the radiation cannot penetrate into the abnormal media as shown in the figure.
Also, the energy density behind the absorber becomes lower in the cases with higher absorption coefficients. 
However, the shadow region is filled with photons due to the scattering processes as the distance from the absorber increases.

\begin{figure}
	\begin{center}
		\includegraphics[width=10.0 cm,clip]{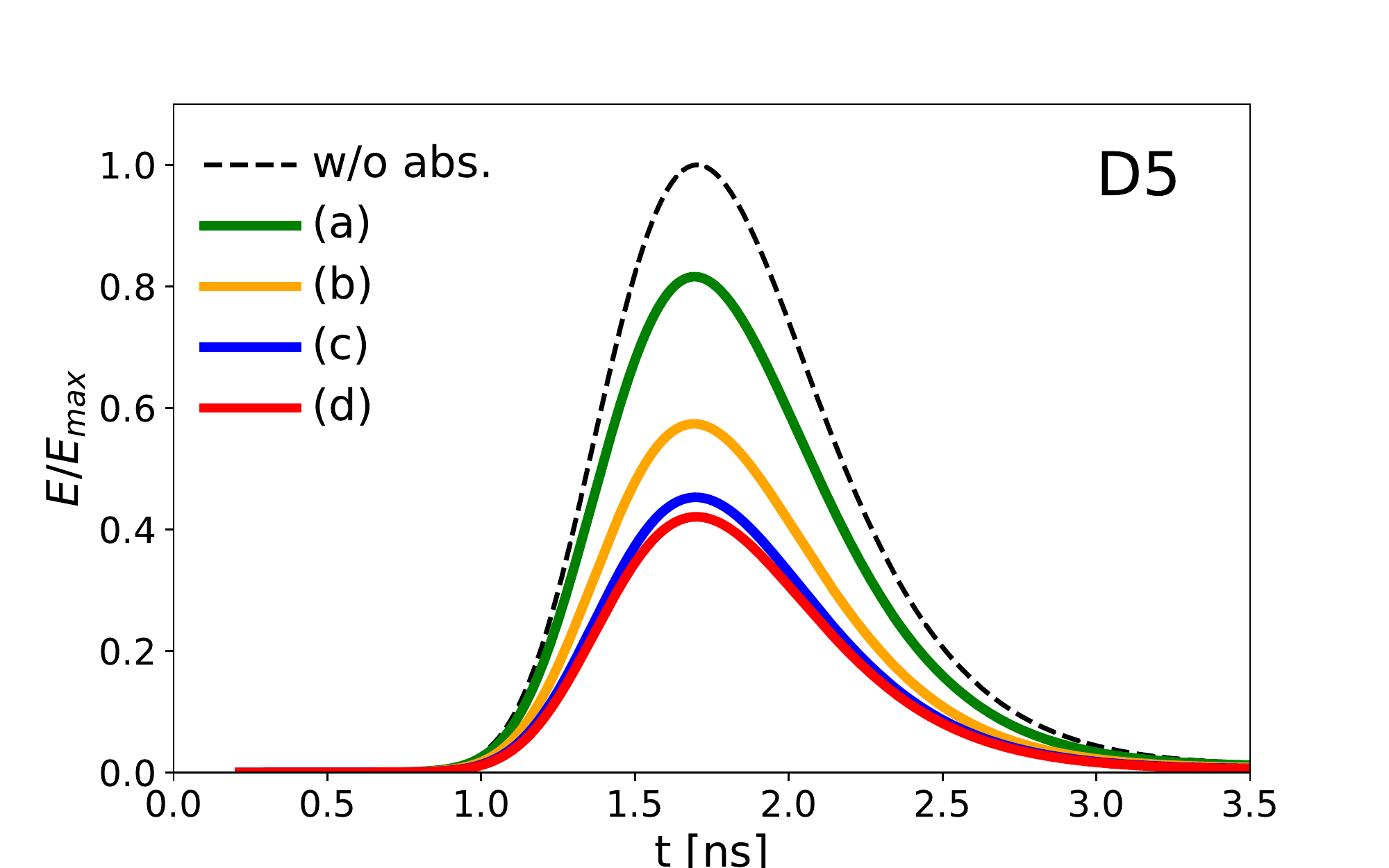}
	\end{center}
	\vspace{-5mm}
	\caption{
		Time-resolved light signals at the detector D5. Different lines are corresponding to the cases shown in Figure~\ref{fig:2dmapwabs}: green (a), yellow (b), blue (c), and red (d). Black dashed line represents the case without the higher absorption area. The normalization factor $E_{\rm max}$ is taken from the peak value of the black dashed line. 
		}  
	\label{fig:out_wabs}
\end{figure}

The time-domain signals at the detector D5 are presented in Figure~\ref{fig:out_wabs}.
Compared to the signals in the case without an absorber, the signals are weakened as the absorption coefficient becomes larger. However, once the optical depth of the absorber exceeds unity significantly, the reduction rate of the signal becomes almost constant. The signals of the case (c) and (d) are lower than the case without the absorber by a factor of $\sim 2$.  
Note that, the arrival times of the peak signals are the same ($t \sim 1.7~\rm ns$) irrespective of the absorption coefficients. 
The signal could also change with the position of the absorber. 
Therefore, that allows us to investigate the properties of an absorber in biological tissue by confronting signals in simulations with those in experiments.

In the above test simulations, we have studied the light propagations in the cases with the cylinder shape absorbers. Here, we additionally investigate the light propagation in the phantom with multiple spheres with a high absorption coefficient. 
We put five spheres of stronger absorption  with a radius of $0.5~\rm cm$ at $[x(\rm cm),y(\rm cm),z(\rm cm)]=[2.0, 1.6, 2.0], [1.5, 2.6, 1.0], [1.0, 0.6, 2.5], [2.3, 1.4, 2.3]$ 
and $[3.0, 2.2, 2.8]$. Note that, two absorbers at $[x(\rm cm),y(\rm cm),z(\rm cm)]=[2.0, 1.6, 2.0]$ and $[2.3, 1.4, 2.3]$ are connected. The absorption efficiency is increased from background media by a factor of 100. The spatial distributions of the absorption coefficient are inhomogeneous and not symmetric to any direction. 

\begin{figure}
	\begin{center}
		\includegraphics[width=7.0 cm,clip]{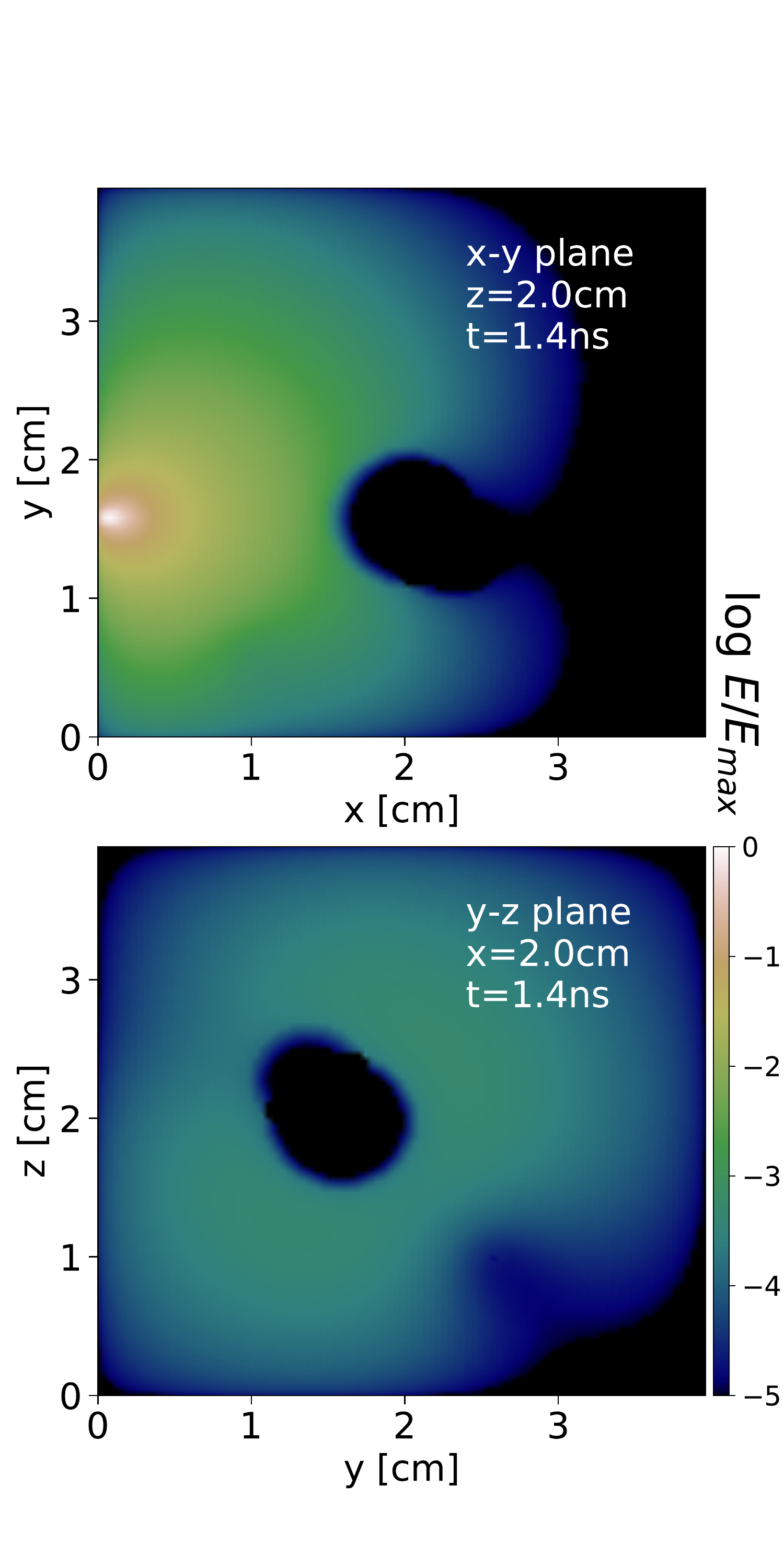}
	\end{center}
	\vspace{-5mm}
	\caption{
	Two-dimensional photon energy distributions for a snapshot at $t=1.4~\rm ns$. 
	Upper and lower panels show $x-z$ and $x-y$ planes at $y=2.0$ and $z=2.0~\rm cm$, respectively.
	Five spheres of stronger absorption  with a radius of $0.5~\rm cm$ are set at
	$[x(\rm cm),y(\rm cm),z(\rm cm)]=[2.0, 1.6, 2.0], [1.5, 2.6, 1.0], [1.0, 0.6, 2.5], [2.3, 1.4, 2.3]$ and $[3.0, 2.2, 2.8]$.  The absorption coefficient is increased from background media by a factor of 100. 
	The energy density is normalized by the peak value of upper panel.
		}
	\label{fig:2dmap5abs}
\end{figure}

Figure~\ref{fig:2dmap5abs} shows two-dimensional photon energy distributions for a snapshot at $t=1.4~\rm ns$. In the panel of $x-y$ plane at $z=2.0~\rm cm$, a shadow due to the connecting two absorbers is seen. Since the absorption optical depth over an absorption sphere is $22.0$, most photons cannot penetrate it. 
In the panel of $y-z$ plane at $x=2.0~\rm cm$, in addition to the shadow of the connecting two absorbers, a shadow due to the absorber at $[x(\rm cm),y(\rm cm),z(\rm cm)]=[1.5, 2.6, 1.0]$ is also seen. Since the center position of the absorber is far from the $y-z$ plane, the contrast of shadow is lower because of the diffuse photons. 
Thus, we confirm that TRINITY can handle radiative transfer problems in highly inhomogeneous media. 

\subsection{Phantom test for a medium with a different refractive index}
The refractive index of biological tissues is higher than that of the vacuum or air by a factor $\sim 1.5$.
Therefore, if there is a part including air like a trachea, photons can be scattered at the boundary, based on Equation~(\ref{eq:reflection}). 
In addition, the light speed changes depending on the local refractive index.  Fujii et al. (2017)\cite{Fujii2017} took into account these effects in two-dimensional simulations and showed that organs with different optical properties change the light signals significantly. 
To perform three-dimensional simulations including these effects, we 
perform simulations with TRINITY. 
First, we regard nearest grids to a region with different refractive index as the boundary grids. In calculations of $\sr$ at the boundary grids, we use $p({\bf r}, \oomega', \oomega)$ taking Equation (\ref{eq:refraction}) and (\ref{eq:reflection}) into account. 

\begin{figure}
	\begin{center}
		\includegraphics[width=10.0 cm,clip]{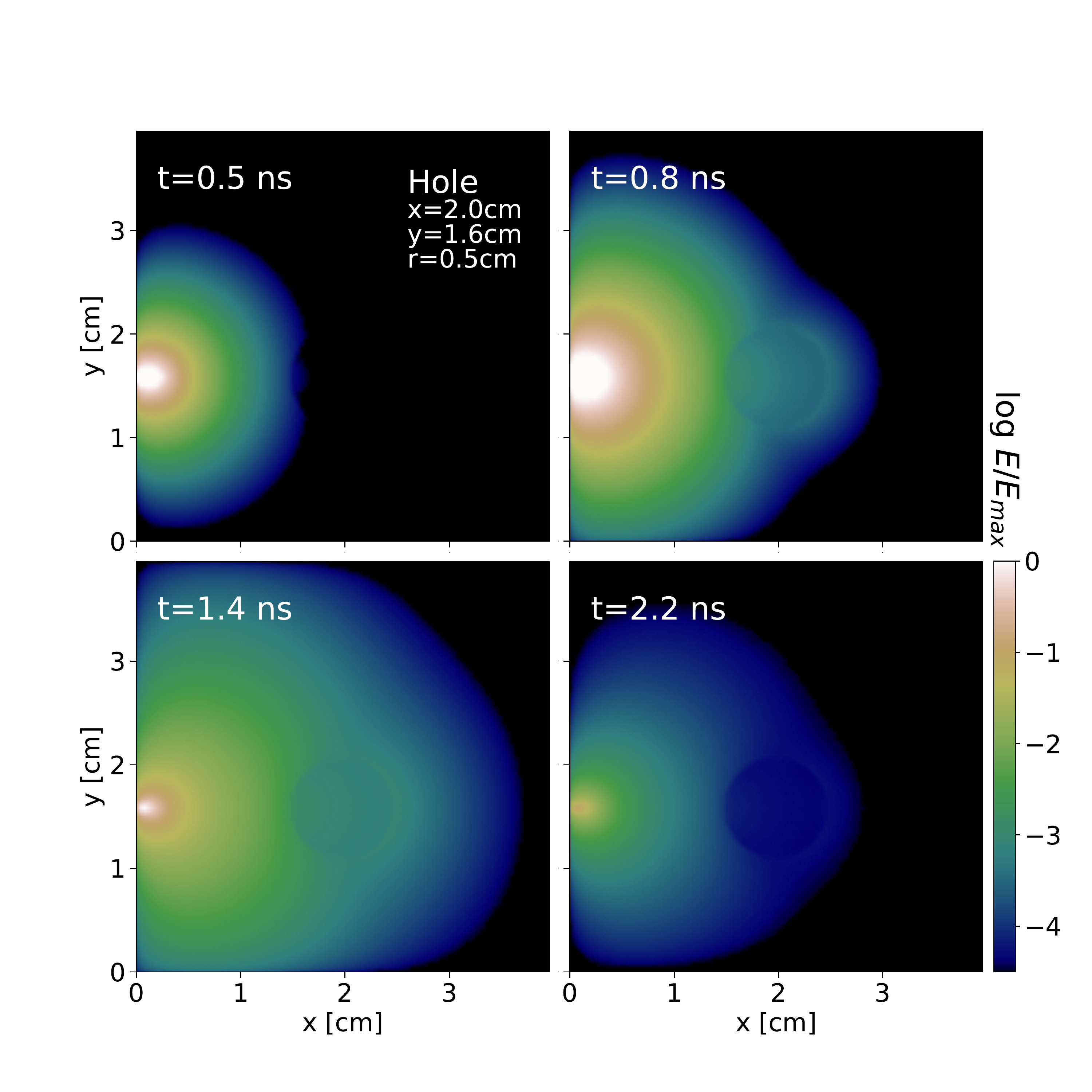}
	\end{center}
	\vspace{-5mm}
	\caption{
	Same as figure~\ref{fig:2dmap}, but with a vacuum hole at $x=2.0$ cm and $y=1.6$ cm with a radius of $0.5~\rm cm$.  
	The energy density is  normalized by the peak value at $t=1.4~\rm ns$.
		}  
	\label{fig:2dmap_neck}
\end{figure}

Actually, we calculate the photon propagation in a phantom with a cylinder hole filled with air. The cylinder hole is located at $x=1.6~\rm cm$ and $y=2.0~\rm cm$ and the radius is $0.5~\rm cm$.  
Figure~\ref{fig:2dmap_neck} presents the energy distributions in $x-y$ plane with $z=2.0~\rm cm$
at $t=0.5, 0.8, 1.4$ and $2.2$ ns. 
At $t \sim 1.4~\rm ns$, the wavefront of the energy peak arrives at the hole. 
Owing to the higher speed of light, the wavefront propagates faster in the hole than other places. 
After the pulse injection is turned off, the photons are trapped for a while due to the scattering. 
However, because of no scattering in the hole, the photon energy density decreases rapidly. 

\begin{figure}
	\begin{center}
		\includegraphics[width=10.0 cm,clip]{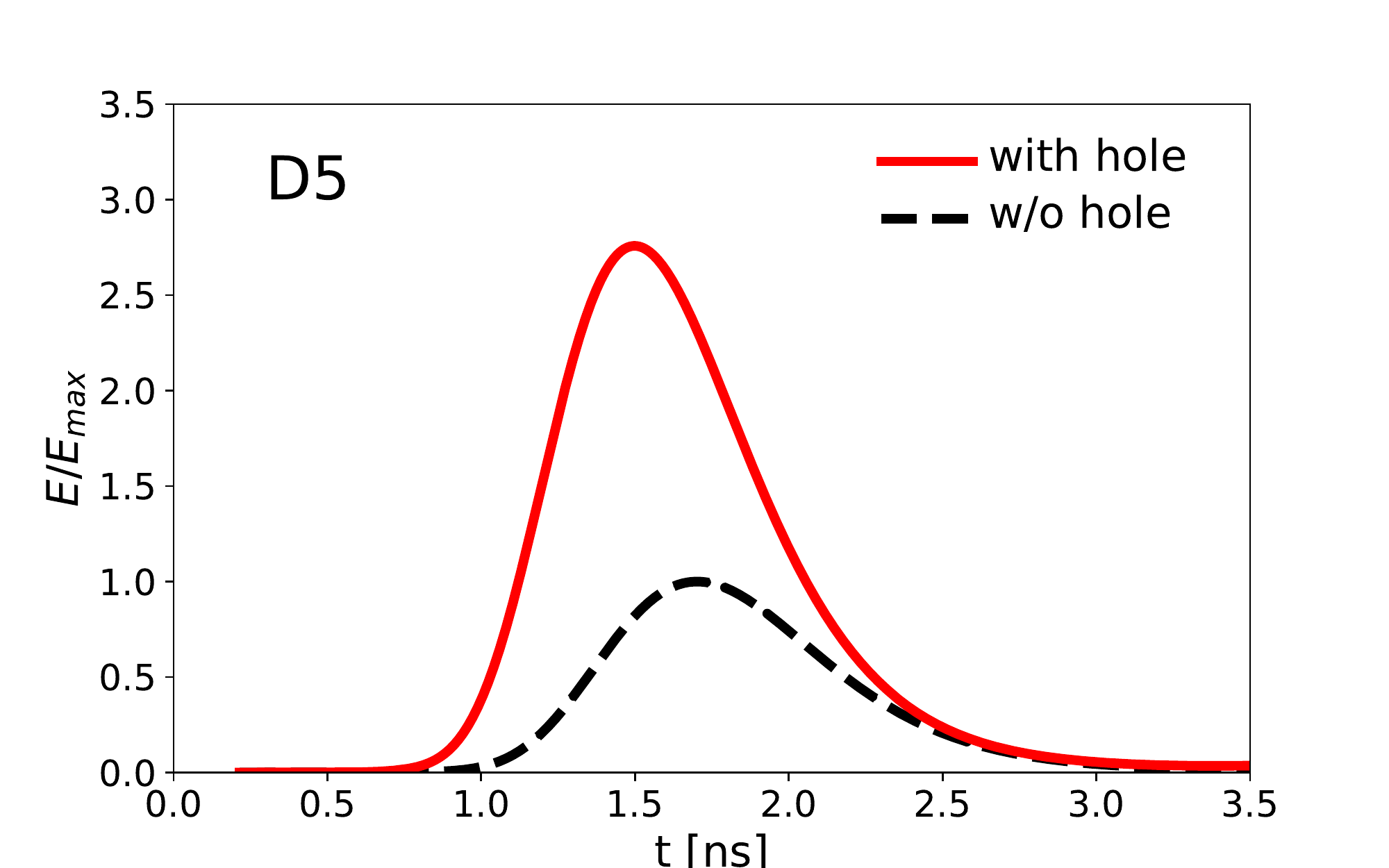}
	\end{center}
	\vspace{-5mm}
	\caption{
	Same as figure~\ref{fig:out_wabs}, but for the case with a vacuum hole with the radius of 0.5 cm. 
	Red solid and black dashed lines represent the results with and without the vacuum hole. The values are normalized by the peak energy of black dashed line.
	}  
	\label{fig:neck_out}
\end{figure}

We present the time-domain signals at the detector D5 in Figure~\ref{fig:neck_out}.
The arrival time of the energy peak is shifted to $\sim 0.2~\rm ns$ earlier because of the faster speed of light in the hole. In addition, the energy peak is higher than that without the hole by a factor of $\sim 3$ because of no absorption there. 
Thus, the light signal significantly changes in the medium with the different refractive index.  
For example, if we consider a neck, the above processes should be taken into account for an accurate diagnosis. 
Note that, it is difficult to handle the refraction and reflection in DE simulations. 
Therefore, RTE simulations are required to properly model the light propagation in complicated parts.

\subsection{MPI parallelization efficiency}

 The above simulations have been performed on the Oakforest-PACS Supercomputer System developed by the University of Tokyo and University of Tsukuba. 
 The Oakforest-PACS has 8208 calculation nodes and its total theoretical calculating performance is 25.0 petaflops. Each calculation node has a CPU, Intel Xeon Phi 7250 that has 68 calculation cores. 
We have used 64 nodes (4352 cores) for a three-dimensional radiative transfer simulation. Each run takes $\sim 20$ hours. 
Here, we investigate MPI parallelization efficiency by changing the number of calculation nodes. 
As in Section 3.1, we consider the phantom with the size of $4.0 \times 4.0 \times 4.0~\rm cm$ and the pulse injection.

\begin{figure}
	\begin{center}
		\includegraphics[width=8.0 cm,clip]{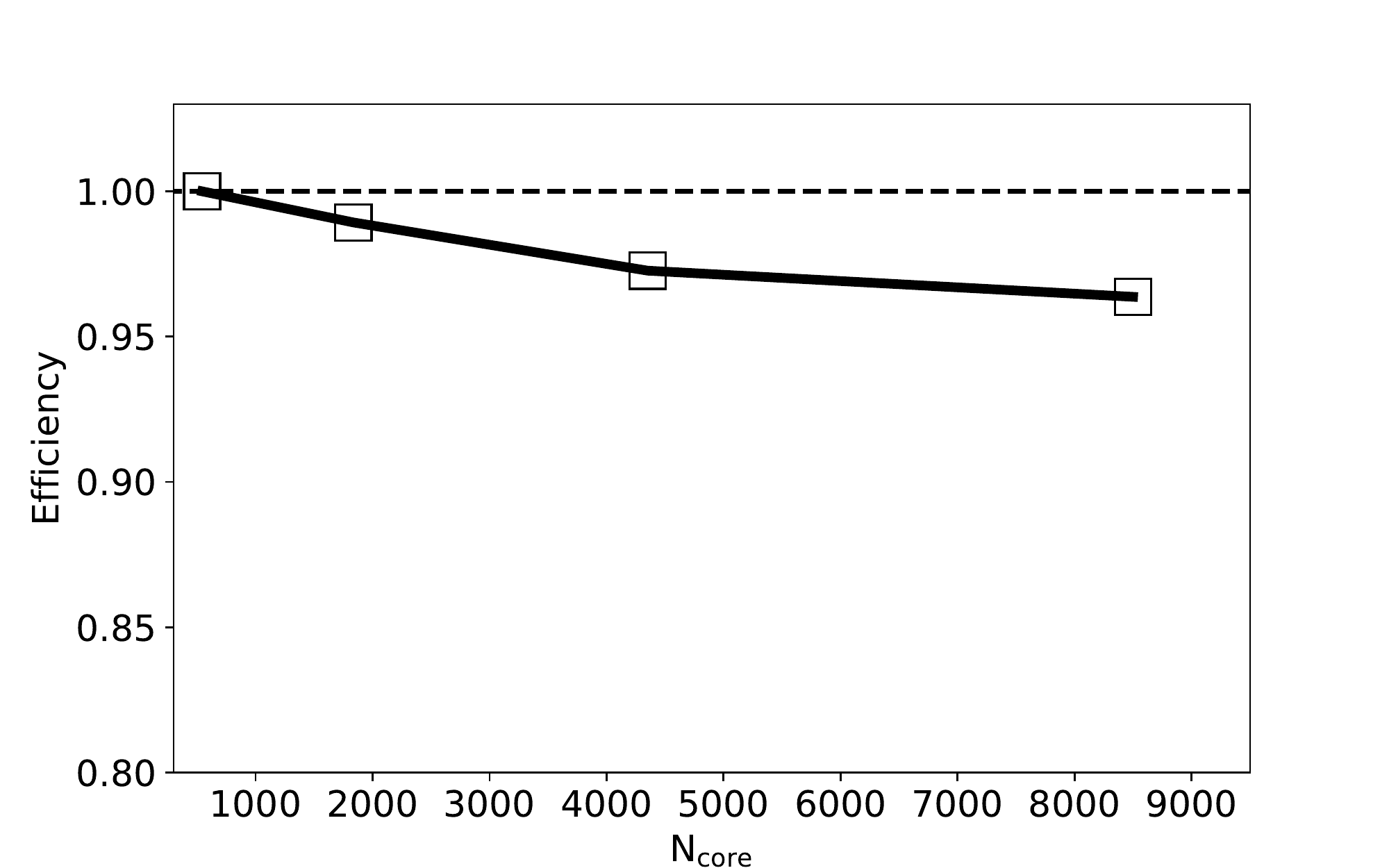}
	\end{center}
	\vspace{-5mm}
	\caption{
	MPI parallelization efficiency as a function of the number of CPU cores.
	The calculation time in the case of 544 cores is chosen as a standard. 
	The efficiency  is measured as $\Delta t_{544} / \Delta t$, where $\Delta t_{544}$ and $\Delta t$ are the calculation times of each time step for the case of 544 cores and changing number of cores.  
	 }
	\label{fig:mpi_eff}
\end{figure}

Figure~\ref{fig:mpi_eff} shows the MPI parallelization efficiency as a function of the number of CPU cores used in a simulation.  Depending on the number of CPU cores, we change the number of spacial grids as 
$N_{\rm grid} = 41^{3}$ for 544 cores (8 nodes), $N_{\rm grid} = 61^{3}$ for 1836 cores (27 nodes), 
$N_{\rm grid} = 81^{3}$ for 4352 cores (64 nodes) and $N_{\rm grid} = 101^{3}$ for 8500 cores (125 nodes).
In this parallelization test, since the calculation amount on each node (each rank) is almost the same, 
the calculation time of each time step should be similar if the MPI parallelization efficiency is high. 
Here, by using the calculation time in the case of 544 cores as a standard, we measure the efficiency 
as  $\Delta t_{544} / \Delta t$, where $\Delta t_{544}$ and $\Delta t$ are the calculation times of each time step for the cases of 544 cores and changing number of cores. 
We find that the simulations keep the high efficiencies of $> 0.95$. Even for the case of 8500 cores, the time for data communication between different domains (CPUs) is much shorter than that in radiative transfer calculations. 
Therefore, even if a large part of a body would be targeted, the simulations would be performed within a reasonable time by utilizing supercomputers with a lot of CPUs.  
 
%
%


%
%

\section{Discussion \& Summary}
Diffuse optical tomography (DOT) is a potential tool to diagnose biological tissues and has several advantages as no exposure, no invasive, and no side effect for instance. 
In order to reconstruct images of biological tissues, theoretical modeling 
based on time-dependent RTE is demanded. 
However, the radiative transfer calculations in three-dimensional space is expensive even with the current computational facilities.
Hence,  the methodology based on RTE has not been established hitherto. 
In this study, to realize three-dimensional radiative transfer calculations with high-spatial resolution, 
we have developed a novel three-dimensional time-dependent radiative transfer code, TRINITY (Time-dependent Radiative transfer In Near-Infrared TomographY), 
for in-vivo DOT. 

We have shown that TRINITY can calculate the light propagation with little numerical diffusion. 
In addition, by utilizing MPI and OpenMP parallelization, we have performed the radiative transfer simulations with spatial grids of $101^{3}$ and angle bins of $3072$, which allow us to resolve a part of the human body with the spatial resolution of $\sim 1~\rm mm$. 

As a first application, we have studied the light propagation in a uniform media made of polyurethane for a pulse with a short irradiation time of  a few nanoseconds. 
We have shown that the initially collimated pulse within $\theta \sim 15^\circ$ spreads due to the multiple scattering processes in all forward directions over several millimeters.
It is demonstrated that
our simulations successfully reproduce the measurements of the time-resolved signals at eight detectors. 
Also, we have investigated cases with an embedded absorber with higher absorption coefficients. In these cases, photons disappear in the absorber and the shadow regions emerge behind it. The simulations show that the shadow regions are gradually recovered due to the scattering processes. Besides,  we have introduced the refraction and reflection effects at boundaries between media with different refractive indices. We have demonstrated the faster photon propagation in an air hole surrounded in the box of polyurethane with the refractive index of $1.51$. 

In this study, we have focused on the methodology of time-dependent radiation
transfer code and the applications to idealized cases. 
By combining the radiative transfer simulations and the inverse problem analysis with the machine-learning \citep{Takamizu2020}, the structure related to the optical properties in a body can be evaluated.
We will investigate the light propagation in a more complicated structure as a human brain or neck that include trachea, blood vessel, spine and spinal cord in future work. 

%
%
\section*{Acknowledgments}

The numerical simulations were performed on the computer clusters at Center for Computational Sciences in University of Tsukuba and Oakforest-PACS at University of Tokyo.
This work is supported in part by Department of Computational Medical Science,
Center for Computational Sciences, University of Tsukuba, the MEXT/JSPS KAKENHI Grant Number 17H04827, 20H04724, 21H04489, National Astronomical Observatory of Japan (NAOJ) ALMA Scientific Research Grant Number 2019-11A,  JST FOREST Program, Grant Number JPMJFR202Z (HY), Japan Agency for Medical Research and Development (AMED)16im0402003h006 (YH). 




\end{document}